\documentclass[nofootinbib,preprint]{revtex4}%
\usepackage{hyperref}
\usepackage{amsmath}
\usepackage{amsfonts}
\usepackage{amssymb}
\usepackage{graphicx}%
\begin{document}
\title{Scatterings of Massive String States from D-brane and \ Their Linear Relations
at High Energies}
\author{Chuan-Tsung Chan}
\email{ctchan@thu.edu.tw}
\affiliation{Department of Physics, Tung-Hai University, Taichung, Taiwan, R.O.C.}
\author{Jen-Chi Lee}
\email{jcclee@cc.nctu.edu.tw}
\affiliation{Department of Electrophysics, National Chiao-Tung University, Hsinchu, Taiwan, R.O.C.}
\author{Yi Yang}
\email{yyang@phys.cts.nthu.edu.tw}
\affiliation{Department of Electrophysics, National Chiao-Tung University and Physics
Division, National Center for Theoretical Sciences, Hsinchu, Taiwan, R.O.C.}
\date{\today}

\begin{abstract}
We study scatterings of bosonic massive closed string states at arbitrary mass
levels from D-brane. We discover that all the scattering amplitudes can be
expressed in terms of the generalized hypergeometric function $_{3}F_{2}$ with
special arguments, which terminates to a finite sum and, as a result, the
whole scattering amplitudes consistently reduce to the usual beta function.
For the simple case of D-particle, we explicitly calculate high-energy limits
of a series of the above scattering amplitudes for arbitrary mass levels, and
derive infinite linear relations among them for each fixed mass level. The
ratios of these high-energy scattering amplitudes are found to be consistent
with the decoupling of high-energy zero-norm states of our previous works.

\end{abstract}
\maketitle

\section{Introduction}

In contrast to the scattering of massless string states, massive higher-spin
string scattering amplitudes \cite{Massive} had not been well studied in the
literature since the developement of string theory. Recently high-energy,
fixed angle behavior of string scattering amplitudes \cite{GM, Gross,
GrossManes} was intensively reinvestigated for massive string states at
arbitrary mass levels \cite{ChanLee1,ChanLee2,
CHL,CHLTY,PRL,paperB,susy,Closed,HL}. The motivation was to uncover the
fundamental hidden stringy spacetime symmetry. An important new ingredient of
this approach is the zero-norm states (ZNS) \cite{Massive,ZNS1,ZNS3,ZNS2} in
the old covariant first quantized (OCFQ) string spectrum. One utilizes the
decoupling of ZNS to obtain nonlinear relations or on-shell Ward identities
among string scattering amplitudes. In the high energy limit, many
simplications occur and one can derive linear relations among high-energy
scattering amplitudes of different string states at each fixed but arbitrary
mass levels. Moreover, these linear relations can be used to fix the ratios
among high energy scattering amplitudes of different string states at each
fixed mass level algebraically. This explicitly shows that there is only one
independent component of high-energy scattering amplitude at each mass level.
On the other hand, a saddle-point method was also developed to calculate the
general formula of tree-level high-energy scattering amplitudes of four
arbitrary string states to verify the ratios calculated above. This general
formula expresses all high-energy string scattering amplitudes in terms of
that of four tachyon as conjectured by Gross in 1988 \cite{Gross}.

In this paper, we study scatterings of bosonic massive closed string states at
arbitrary mass levels from D-brane. The scattering of massless string states
from D-brane was well studied in the literature and can be found in
\cite{Klebanov}. Since the mass of D-brane scales as the inverse of the string
coupling constant $1/g$, we will assume that it is infinitely heavy to leading
order in $g$ and does not recoil. We will first show that, for the
$(0\rightarrow1)$ and $(1\rightarrow\infty)$ channels, all the scattering
amplitudes can be expressed in terms of the beta functions, thanks to the
momentum conservation on the D-brane. Alternatively, the Kummer relation of
the hypergeometric function $_{2}F_{1}$ can be used to reduce the scattering
amplitudes to the usual beta function. After summing up the $(0\rightarrow1)$
and $(1\rightarrow\infty)$ channels, we discover that all the scattering
amplitudes can be expressed in terms of the generalized hypergeometric
function $_{3}F_{2}$ with special arguments, which terminates to a finite sum
and, as a result, the whole scattering amplitudes consistently reduce to the
usual beta function. Finally, for the simple case of D-particle, we explicitly
calculate high-energy limit of a series of the above scattering amplitudes for
arbitrary mass levels, and derive infinite linear relations among them for
each fixed mass level. Since the calculation of decoupling of high energy ZNS
remains the same as the case of scatterings without D-brane, the ratios of
these high-energy scattering amplitudes are found to be consistent with the
decoupling of high-energy zero-norm states of our previous works.
\cite{ChanLee1,ChanLee2, CHL,CHLTY,PRL,paperB,susy,Closed}. This paper is
organized as follows. In section II, we first calculate closed string tachyon
scatters to D-brane. The calculation is then generalized to arbitrary mass
levels. By using the Kummer relation of the hypergeometric function $_{2}%
F_{1}$, all the scattering amplitudes in the $(0\rightarrow1)$ and
$(1\rightarrow\infty)$ channels can be reduced to the usual beta function. We
then sum up the $(0\rightarrow1)$ and $(1\rightarrow\infty)$ channels, and
show that all the scattering amplitudes can be expressed in terms of the
terminated generalized hypergeometric function $_{3}F_{2}$ with special
arguments, and the whole scattering amplitudes consistently reduce to the
usual beta function. In section III, for the simple case of D-particle, we
explicitly calculate high-energy limit of a series of the above scattering
amplitudes for arbitrary mass levels, and derive the linear relations among
them. The results are compared to the calculations of decoupling of high
energy ZNS. We give a brief conclusion in section IV. Some relevant formulas
of the hypergeometric function $_{2}F_{1}$ and the generalized hypergeometric
function $_{3}F_{2}$ are listed in the Appendix.%

\setcounter{equation}{0}
\renewcommand{\theequation}{\arabic{section}.\arabic{equation}}%

\section{Scatterings of Massive String States from D-brane}

We first review some massive string scattering amplitudes for arbitrary mass
levels without D-brane. The $(s,t)$ channel scattering amplitude of 26D open
bosonic string with $V_{2}=\alpha_{-1}^{\mu_{1}}\alpha_{-1}^{\mu_{2}}%
\cdot\cdot\cdot\alpha_{-1}^{\mu_{n}}\left\vert 0,k\right\rangle $, the highest
spin state at mass level $M_{op}^{2}$ $=2(n-1)$, and three tachyons
$V_{1,3,4}$ was calculated to be \cite{Closed}%
\begin{equation}
\mathcal{T}_{n;st}^{\mu_{1}\mu_{2}\cdot\cdot\mu_{n}}=\overset{n}%
{\underset{l=0}{\sum}}(-)^{l}\binom{n}{l}B\left(  -\frac{s}{2}-1+l,-\frac
{t}{2}-1+n-l\right)  k_{1}^{(\mu_{1}}..k_{1}^{\mu_{n-l}}k_{3}^{\mu_{n-l+1}%
}..k_{3}^{\mu_{n})},
\end{equation}
where $B(u,v)=\int_{0}^{1}dxx^{u-1}(1-x)^{v-1}$ is the Euler beta function.
The corresponding $(t,u)$ channel scattering amplitude can be calculated to be%
\begin{equation}
\mathcal{T}_{n;tu}^{\mu_{1}\mu_{2}\cdot\cdot\mu_{n}}=\overset{n}%
{\underset{l=0}{\sum}}\binom{n}{l}B\left(  -\frac{t}{2}+n-l-1,-\frac{u}%
{2}-1\right)  k_{1}^{(\mu_{1}}..k_{1}^{\mu_{n-l}}k_{3}^{\mu_{n-l+1}}%
..k_{3}^{\mu_{n})}.\label{ntu}%
\end{equation}
In calculating Eq.(\ref{ntu}), we have used the Mobious transformation
$y=\left(  x-1\right)  /x$ to change the integration region from $\left(
1\rightarrow\infty\right)  $ to $\left(  0\rightarrow1\right)  $. One can see
that all the scattering amplitudes above can be expressed in terms of the beta function.

In this section we will study the general structure of an arbitrary incoming
closed string state scatters from D-brane and ends up with an arbitrary spin
outgoing closed string states at arbitrary mass levels. In particular, we will
examine whether they can be expressed in terms of beta function as the above
scattering amplitudes without D-brane. We will first begin with the simple
case of tachyon to tachyon scattering and then generalize to scatterings of
states at arbitrary mass levels. The standard propagators of the left and
right moving fields are%
\begin{align}
\left\langle X^{\mu}\left(  z\right)  X^{\nu}\left(  w\right)  \right\rangle
&  =-\eta^{\mu\nu}\log\left(  z-w\right)  ,\label{D1}\\
\left\langle \tilde{X}^{\mu}\left(  \bar{z}\right)  \tilde{X}^{\nu}\left(
\bar{w}\right)  \right\rangle  &  =-\eta^{\mu\nu}\log\left(  \bar{z}-\bar
{w}\right)  . \label{D2}%
\end{align}
In addition, there are also nontrivial correlator between the right and left
moving fields as well%
\begin{equation}
\left\langle X^{\mu}\left(  z\right)  \tilde{X}^{\nu}\left(  \bar{w}\right)
\right\rangle =-D^{\mu\nu}\log\left(  z-\bar{w}\right)  \label{DD}%
\end{equation}
as a result of the boundary condition at the real axis. Propagator
Eq.(\ref{DD}) has the standard form Eq.(\ref{D1}) for the fields satisfying
Neumann boundary condition, while matrix $D$ reverses the sign for the fields
satisfying Dirichlet boundary condition. We will follow the standard notation
and make the following replacement%
\begin{equation}
\tilde{X}^{\mu}\left(  \bar{z}\right)  \rightarrow D_{\text{ \ }\nu}^{\mu
}X^{\nu}\left(  \bar{z}\right)
\end{equation}
which allows us to use the standard correlators Eq.(\ref{D1}) throughout our
calculations. As we will see, the existence of the Propagator Eq.(\ref{DD})
has far-reaching effect on the string scatterings from D-brane.

\subsection{Tachyon to tachyon}

We first consider the tachyon to tachyon scattering amplitude%

\begin{align}
A_{tach} &  =\int d^{2}z_{1}d^{2}z_{2}\left\langle V_{1}\left(  z_{1},\bar
{z}_{1}\right)  V_{2}\left(  z_{2},\bar{z}_{2}\right)  \right\rangle
\nonumber\\
&  =\int d^{2}z_{1}d^{2}z_{2}\left\langle V\left(  k_{1},z_{1}\right)
\tilde{V}\left(  k_{1},\bar{z}_{1}\right)  V\left(  k_{2},z_{2}\right)
\tilde{V}\left(  k_{2},\bar{z}_{2}\right)  \right\rangle \nonumber\\
&  =\int d^{2}z_{1}d^{2}z_{2}\left\langle e^{ik_{1}X\left(  z_{1}\right)
}e^{ik_{1}\tilde{X}\left(  \bar{z}_{1}\right)  }e^{ik_{2}X\left(
z_{2}\right)  }e^{ik_{2}\tilde{X}\left(  \bar{z}_{2}\right)  }\right\rangle
\nonumber\\
&  =\int d^{2}z_{1}d^{2}z_{2}\left(  z_{1}-\bar{z}_{1}\right)  ^{k_{1}\cdot
D\cdot k_{1}}\left(  z_{2}-\bar{z}_{2}\right)  ^{k_{2}\cdot D\cdot k_{2}%
}\left\vert z_{1}-z_{2}\right\vert ^{2k_{1}\cdot k_{2}}\left\vert z_{1}%
-\bar{z}_{2}\right\vert ^{2k_{1}\cdot D\cdot k_{2}}.
\end{align}
To fix the $SL\left(  2,R\right)  $ invariance, we set $z_{1}=iy$ and
$z_{2}=i$. Introducing the $SL\left(  2,R\right)  $ Jacobian%
\begin{equation}
d^{2}z_{1}d^{2}z_{2}=4\left(  1-y^{2}\right)  dy,
\end{equation}
we have, for the $(0\rightarrow1)$ channel,
\begin{align}
A_{tach}^{(0\rightarrow1)} &  =4\left(  2i\right)  ^{k_{1}\cdot D\cdot
k_{1}+k_{2}\cdot D\cdot k_{2}}\int_{0}^{1}dy\text{ }y^{k_{2}\cdot D\cdot
k_{2}}\left(  1-y\right)  ^{2k_{1}\cdot k_{2}+1}\left(  1+y\right)
^{2k_{1}\cdot D\cdot k_{2}+1}\nonumber\\
&  =4\left(  2i\right)  ^{2a_{0}}\int_{0}^{1}dy\text{ }y^{a_{0}}\left(
1-y\right)  ^{b_{0}}\left(  1+y\right)  ^{c_{0}}\nonumber\\
&  =4\left(  2i\right)  ^{2a_{0}}\frac{\Gamma\left(  a_{0}+1\right)
\Gamma\left(  b_{0}+1\right)  }{\Gamma\left(  a_{0}+b_{0}+2\right)  }\text{
}_{2}F_{1}(-c_{0},a_{0}+1,a_{0}+b_{0}+2,-1)\label{ta3}\\
&  =4\left(  2i\right)  ^{2a_{0}}\frac{\Gamma\left(  a_{0}+1\right)
\Gamma\left(  b_{0}+1\right)  }{\Gamma\left(  a_{0}+b_{0}+2\right)
}2^{-2a_{0}-1-N}\text{ }_{2}F_{1}(N-a_{0},b_{0}+1,a_{0}+b_{0}+2,-1)\nonumber\\
&  =4\left(  2i\right)  ^{2a_{0}}2^{-2a_{0}-1-N}\int_{0}^{1}dt\text{ }%
t^{b_{0}}\left(  1-t\right)  ^{a_{0}}\left(  1+t\right)  ^{a_{0}%
+N}.\label{ta5}%
\end{align}
In the above calculations, we have defined
\begin{align}
a_{0} &  =k_{1}\cdot D\cdot k_{1}=k_{2}\cdot D\cdot k_{2},\\
b_{0} &  =2k_{1}\cdot k_{2}+1,\\
c_{0} &  =2k_{1}\cdot D\cdot k_{2}+1,
\end{align}
so that%
\begin{equation}
2a_{0}+b_{0}+c_{0}+2=4N_{1}\equiv-N,
\end{equation}
and $-k_{1}^{2}=M^{2}\equiv\frac{M_{closed}^{2}}{2\alpha_{closed}^{\prime}%
}=2(N_{1}-1)$, $N_{1}=0$ for tachyon. We have also used the integral
representation of the hypergeometric function
\begin{equation}
_{2}F_{1}\left(  \alpha,\beta,\gamma,z\right)  =\frac{\Gamma\left(
\gamma\right)  }{\Gamma\left(  \beta\right)  \Gamma\left(  \gamma
-\beta\right)  }\int_{0}^{1}dt\text{ }t^{\beta-1}\left(  1-t\right)
^{\gamma-\beta-1}\left(  1-zt\right)  ^{-\alpha},\label{gama}%
\end{equation}
and the following identity%
\begin{equation}
_{2}F_{1}(\alpha,\beta,\gamma;x)=(1-x)^{\gamma-\alpha-\beta}\text{ }_{2}%
F_{1}(\gamma-\alpha,\gamma-\beta,\gamma;x),\label{half}%
\end{equation}
which we discuss in the appendix. In addition, the momentum conservation on
the D-brane%
\begin{equation}
D\cdot k_{1}+k_{1}+D\cdot k_{2}+k_{2}=0\label{con}%
\end{equation}
is crucial to get the final result Eq.(\ref{ta5}).\ Finally, by using change
of variable $\tilde{t}=t^{2}$, Eq.(\ref{ta5}) can be further reduced to the
beta function%
\begin{equation}
A_{tach}^{(0\rightarrow1)}\simeq\frac{\Gamma\left(  a_{0}+1\right)
\Gamma\left(  \frac{b_{0}+1}{2}\right)  }{\Gamma\left(  a_{0}+\frac{b_{0}}%
{2}+\frac{3}{2}\right)  }=B\left(  a_{0}+1,\frac{b_{0}+1}{2}\right)
\label{final}%
\end{equation}
where we have omitted an irrelevant factor.

For the $(1\rightarrow\infty)$ channel, we use the change of variable
$y=\frac{1+t}{1-t}$ and end up with the same result%
\begin{align}
A_{tach}^{(1\rightarrow\infty)}  &  =4\left(  2i\right)  ^{k_{1}\cdot D\cdot
k_{1}+k_{2}\cdot D\cdot k_{2}}\int_{1}^{\infty}dy\text{ }y^{k_{2}\cdot D\cdot
k_{2}}\left(  y-1\right)  ^{2k_{1}\cdot k_{2}+1}\left(  1+y\right)
^{2k_{1}\cdot D\cdot k_{2}+1}\nonumber\\
&  =4\left(  2i\right)  ^{2a_{0}}2^{-2a_{0}-1-N}\int_{0}^{1}dt\text{ }%
t^{b_{0}}\left(  1-t\right)  ^{a_{0}+N}\left(  1+t\right)  ^{a_{0}}\nonumber\\
&  \simeq\frac{\Gamma\left(  a_{0}+1\right)  \Gamma\left(  \frac{b_{0}+1}%
{2}\right)  }{\Gamma\left(  a_{0}+\frac{b_{0}}{2}+\frac{3}{2}\right)
}=B\left(  a_{0}+1,\frac{b_{0}+1}{2}\right)
\end{align}
since $N=0$ for the case of tachyon.

Altenatively, one can use the Kummer formula of hypergeometric function%
\begin{equation}
_{2}F_{1}(\alpha,\beta,1+\alpha-\beta,-1)=\frac{\Gamma(1+\alpha-\beta
)\Gamma(1+\frac{\alpha}{2})}{\Gamma(1+\alpha)\Gamma(1+\frac{\alpha}{2}-\beta)}%
\end{equation}
and%
\begin{equation}
\Gamma\left(  \frac{1+\alpha}{2}\right)  =\frac{2^{-\alpha}\sqrt{\pi}%
\Gamma\left(  1+\alpha\right)  }{\Gamma\left(  1+\frac{\alpha}{2}\right)  },
\end{equation}
to reduce Eq.(\ref{ta3}) to the final result Eq.(\ref{final}). In this
calculation, we have used the Kummer condition%
\begin{equation}
\gamma=1+\alpha-\beta,
\end{equation}
which is equivalent to the momentum conservation on the D-brane Eq.(\ref{con}).

\subsection{Tensor to tensor}

\bigskip In this subsection, we generalize the previous calculation to general
tensor to tensor scatterings. In this case, we define%
\begin{align}
a  &  =k_{1}\cdot D\cdot k_{1}+n_{a}\equiv a_{0}+n_{a},\\
b  &  =2k_{1}\cdot k_{2}+1+n_{b}\equiv b_{0}+n_{b},\\
c  &  =2k_{1}\cdot D\cdot k_{2}+1+n_{c}\equiv c_{0}+n_{c},
\end{align}
where $n_{a}$, $n_{b}$ and $n_{c}$ are integer and%
\[
N^{\prime}=-\left(  2n_{a}+n_{b}+n_{c}\right)  ,
\]
so that%
\begin{equation}
2a+b+c+2+N^{\prime}=4N_{1}\Longrightarrow2a+b+c+2=4N_{1}-N^{\prime}\equiv-N
\label{int}%
\end{equation}
where $k_{1}^{2}=2(N_{1}-1)$ and $N_{1}$ is now the mass level of $k_{1}$.
After a similar calculation as the previous subsection, it is easy to see that
a typical term in the expression of the general tensor to tensor scattering
amplitudes can be reduced to the following integral%

\begin{align}
I_{\left(  0\rightarrow1\right)  }  &  =\int_{0}^{1}dt\text{ }t^{a}\left(
1-t\right)  ^{b}\left(  1+t\right)  ^{c},\nonumber\\
&  =\frac{\Gamma\left(  a+1\right)  \Gamma\left(  b+1\right)  }{\Gamma\left(
a+b+2\right)  }\text{ }_{2}F_{1}\left(  -c,a+1,a+b+2,-1\right) \nonumber\\
&  =2^{b+c+1}\frac{\Gamma\left(  a+1\right)  \Gamma\left(  b+1\right)
}{\Gamma\left(  a+b+2\right)  }\text{ }_{2}F_{1}\left(
-a-N,b+1,a+b+2,-1\right) \nonumber\\
&  =2^{-2a-1-N}\int_{0}^{1}dt\text{ }t^{b}\left(  1-t\right)  ^{a}\left(
1+t\right)  ^{a+N}.
\end{align}
Similarly, for the $(1\rightarrow\infty)$ channel, one gets
\begin{align}
I_{\left(  1\rightarrow\infty\right)  }  &  =\int_{1}^{\infty}dy\text{ }%
y^{a}\left(  y-1\right)  ^{b}\left(  1+y\right)  ^{c}\nonumber\\
&  =2^{-2a-1-N}\int_{0}^{1}dt\text{ }t^{b}\left(  1-t\right)  ^{a+N}\left(
1+t\right)  ^{a}.
\end{align}
The sum of the two channels gives%
\begin{align}
I  &  =I_{\left(  0\rightarrow1\right)  }+I_{\left(  1\rightarrow
\infty\right)  }\nonumber\\
&  =2^{-2a-1-N}\int_{0}^{1}dt\text{ }t^{b}\left(  1-t\right)  ^{a}\left(
1+t\right)  ^{a}\left[  \left(  1+t\right)  ^{N}+\left(  1-t\right)
^{N}\right] \nonumber\\
&  =2^{-2a-1-N}\sum_{m=0}^{N}\left[  1+\left(  -1\right)  ^{m}\right]
\binom{N}{m}\int_{0}^{1}dt\text{ }t^{b+m}\left(  1-t\right)  ^{a}\left(
1+t\right)  ^{a}\nonumber\\
&  =2^{-2a-2-N}\sum_{m=0}^{N}\left[  1+\left(  -1\right)  ^{m}\right]
\binom{N}{m}\cdot\frac{\Gamma\left(  a+1\right)  \Gamma\left(  \frac{b+1}%
{2}+\frac{m}{2}\right)  }{\Gamma\left(  a+\frac{b+3}{2}+\frac{m}{2}\right)
}\nonumber\\
&  =2^{-2a-1-N}\frac{\Gamma\left(  a+1\right)  \Gamma\left(  \frac{b+1}%
{2}\right)  }{\Gamma\left(  a+\frac{b+3}{2}\right)  }\sum_{n=0}^{\left[
\frac{N}{2}\right]  }\binom{N}{2n}\dfrac{\left(  \frac{b+1}{2}\right)  _{n}%
}{\left(  a+\frac{b+3}{2}\right)  _{n}}\nonumber\\
&  =2^{-2a-1-N}\cdot B\left(  a+1,\frac{b+1}{2}\right)  \cdot\text{ }_{3}%
F_{2}\left(  \frac{b+1}{2},-\left[  \frac{N}{2}\right]  ,\dfrac{1}{2}-\left[
\frac{N}{2}\right]  ;a+\frac{b+3}{2},\dfrac{1}{2};1\right)  \label{master}%
\end{align}
where the generalized hypergeometric function $_{3}F_{2}$ is defined in the
appendix. Note that the energy dependence of the prefactor $4\left(
2i\right)  ^{k_{1}\cdot D\cdot k_{1}+k_{2}\cdot D\cdot k_{2}}$ in the
scattering amplitude cancels, apart from an irrelvant factor, the energy
dependence of $2^{-2a-1-N}$ by using Eq.(\ref{int}). For $N=0$, one recovers
the result of tachyon scattering amplitude Eq.(\ref{ta5}). For the special
arguments of $_{3}F_{2}$ in Eq.(\ref{master}), the hypergeometric function
terminates to a finite sum and, as a result, the whole scattering amplitudes
consistently reduce to the usual beta function. The explicit forms of
$_{3}F_{2}$ for some integer $N$ are given in the appendix.%

\setcounter{equation}{0}
\renewcommand{\theequation}{\arabic{section}.\arabic{equation}}%

\section{High Energy Scattering Amplitudes}

In this section, we will calculate the high energy limit of string scattered
from D-brane. In particular, we will calculate the ratios among scattering
amplitudes of different string states at high energies. We first begin with a
brief review of the calculation of these ratios without D-brane. There are
three methods to calculate these ratios \cite{CHLTY,PRL}. We will only review
the method of decoupling of high-energy ZNS for 26D bosonic open string
theory. The same ratios can be obtained by two other methods, the high energy
Virasoro constraints and the saddle-point method. Since the calculation of
decoupling of high energy ZNS without D-brane remains the same as the
calculation of string scatters from D-brane, the ratios of these high-energy
scattering amplitudes can be used to check the ratios we will obtain for
string scatters from D-brane. At a fixed mass level $M_{op}^{2}=2(n-1)$, it
was shown that \cite{CHLTY,PRL} a four-point function is at the leading order
in high-energy limit only for states of the following form ( we use the
notation of \cite{GSW})
\begin{equation}
\left\vert n,2m,q\right\rangle \equiv(\alpha_{-1}^{T})^{n-2m-2q}(\alpha
_{-1}^{L})^{2m}(\alpha_{-2}^{L})^{q}\left\vert 0,k\right\rangle .\label{1}%
\end{equation}
where $n\geqslant2m+2q,m,q\geqslant0.$The state in Eq.(\ref{1}) is arbitrarily
chosen to be the second vertex of the four-point function. The other three
points can be any string states. We have defined the normalized polarization
vectors of the second string state to be \cite{ChanLee1,ChanLee2}
\begin{equation}
e_{P}=\frac{1}{M_{op}}(E_{2},\mathrm{k}_{2},0)=\frac{k_{2}}{M_{op}},\label{2}%
\end{equation}%
\begin{equation}
e_{L}=\frac{1}{M_{op}}(\mathrm{k}_{2},E_{2},0),\label{3}%
\end{equation}%
\begin{equation}
e_{T}=(0,0,1)\label{4}%
\end{equation}
in the CM frame contained in the plane of scattering. In the OCFQ spectrum of
open bosonic string theory, the solutions of physical states conditions
include positive-norm propagating states and two types of zero-norm states.
The latter are \cite{GSW}%
\begin{equation}
\text{Type I}:L_{-1}\left\vert x\right\rangle ,\text{ where }L_{1}\left\vert
x\right\rangle =L_{2}\left\vert x\right\rangle =0,\text{ }L_{0}\left\vert
x\right\rangle =0;\label{5}%
\end{equation}%
\begin{equation}
\text{Type II}:(L_{-2}+\frac{3}{2}L_{-1}^{2})\left\vert \widetilde
{x}\right\rangle ,\text{ where }L_{1}\left\vert \widetilde{x}\right\rangle
=L_{2}\left\vert \widetilde{x}\right\rangle =0,\text{ }(L_{0}+1)\left\vert
\widetilde{x}\right\rangle =0.\label{6}%
\end{equation}
While Type I states have zero-norm at any space-time dimension, Type II states
have zero-norm \emph{only} at D=26. The decoupling of the following Type I
high-energy zero-norm states (HZNS)%
\begin{equation}
L_{-1}\left\vert n-1,2m-1,q\right\rangle =M_{op}\left\vert n,2m,q\right\rangle
+(2m-1)\left\vert n,2m-2,q+1\right\rangle \label{7}%
\end{equation}
gives the first high-energy Ward identities%
\begin{equation}
\mathcal{T}^{(n,2m,q)}=\left(  -\frac{2m-1}{M_{op}}\right)  \cdots\left(
-\frac{3}{M_{op}}\right)  \left(  -\frac{1}{M_{op}}\right)  \mathcal{T}%
^{(n,0,q+m)}.\label{8}%
\end{equation}
where $\mathcal{T}^{(n,2m,q)}$ represents the four-point functions at level
$n.$ Similarly, the decoupling of the following Type II HZNS%
\begin{equation}
L_{-2}\left\vert n-2,0,q\right\rangle =\frac{1}{2}\left\vert
n,0,q\right\rangle +M_{op}\left\vert n,0,q+1\right\rangle \label{9}%
\end{equation}
gives the second high-energy Ward identities%
\begin{equation}
\mathcal{T}^{(n,0,q)}=\left(  -\frac{1}{2M_{op}}\right)  ^{q}\mathcal{T}%
^{(n,0,0)}.\label{10}%
\end{equation}
Combining Eqs.(\ref{8}) and (\ref{10}) gives the master formula
\cite{CHLTY,PRL,paperB}
\begin{equation}
\mathcal{T}^{(n,2m,q)}=\left(  -\frac{1}{M_{op}}\right)  ^{2m+q}\left(
\frac{1}{2}\right)  ^{m+q}(2m-1)!!\mathcal{T}^{(n,0,0)},\label{11}%
\end{equation}
which shows that there is only one independent high-energy scattering
amplitudes at each fixed mass level. Eq.(\ref{11}) also gives the ratios of
high energy scattering amplitudes among different string states. For the case
of closed string, the ratios are the tensor products of two open string ratios
\cite{Closed}. It is interesting to note that \cite{CHLTY} in calculating the
type II high energy Ward identities Eq.(3.9), we have omitted the second term
$\frac{3}{2}L_{-1}^{2}\left\vert \widetilde{x}\right\rangle $ of type II ZNS
in Eq.(3.6). It turns out that this omition will not, in the high energy
limit, affect the final result.

We now turn to the case of D-brane scatterings. For our purpose here, for
simplicity, we will only consider the case with $m=0$. That is, states in
Eq.(\ref{1}) without $(\alpha_{-1}^{L})^{2m}$component. The reason is as
following. It was shown that \cite{ChanLee1,ChanLee2,CHL} the leading order
amplitudes containing this component will drop from energy order $E^{4m}$ to
$E^{2m}$, and one needs to calculate the complicated naive subleading
contraction terms between $\partial X$ and $\partial X$ for the \textit{multi}%
-tensor scattering in order to get the real leading order scattering
amplitudes. For our \textit{closed} string scattering calculation here, even
for the case of one tachyon and one tensor scattering, one encounters the
similar complicated nonzero contraction terms in Eq.(\ref{DD}) due to the
D-brane. So we will omit high energy scattering amplitudes of string states
containing this $(\alpha_{-1}^{L})^{2m}$component. On the other hand, we will
also need the result that the high energy closed string ratios are the tensor
product of two pieces of open string ratios \cite{Closed}.

To simplify the kinematics, we consider the case of D-0 brane or D-particle
scatterings. The momentum of the incident particle $k_{2}$ is along the $-X$
direction and particle $k_{1}$ is scattered at an angle $\phi$. We will
consider the general case of an incoming tensor state $\left(  \alpha_{-1}%
^{T}\right)  ^{n-2q}\left(  \alpha_{-2}^{L}\right)  ^{q}\otimes\left(
\tilde{\alpha}_{-1}^{T}\right)  ^{n-2q^{\prime}}\left(  \tilde{\alpha}%
_{-2}^{L}\right)  ^{q^{\prime}}\left\vert 0\right\rangle $ and an outgoing
tachyon state. Our result can be easily generalized to the more general two
tensor cases. The kinematic setup is%
\begin{align}
e^{P}  &  =\frac{1}{M}\left(  -E,-\mathrm{k}_{2},0\right)  =\frac{k_{2}}{M},\\
e^{L}  &  =\frac{1}{M}\left(  -\mathrm{k}_{2},-E,0\right)  ,\\
e^{T}  &  =\left(  0,0,1\right)  ,\\
k_{1}  &  =\left(  E,\mathrm{k}_{1}\cos\phi,-\mathrm{k}_{1}\sin\phi\right)
,\\
k_{2}  &  =\left(  -E,-\mathrm{k}_{2},0\right)  .
\end{align}
For the scattering of D-particle $D_{ij}=-\delta_{ij}$, and it is easy to
calculate%
\begin{align}
e^{T}\cdot k_{2}  &  =e^{L}\cdot k_{2}=0,\\
e^{T}\cdot k_{1}  &  =-\mathrm{k}_{1}\sin\phi\sim-E\sin\phi,\\
e^{T}\cdot D\cdot k_{1}  &  =\mathrm{k}_{1}\sin\phi\sim E\sin\phi,\\
e^{T}\cdot D\cdot k_{2}  &  =0,\\
e^{L}\cdot k_{1}  &  =\frac{1}{M}\left[  \mathrm{k}_{2}E-\mathrm{k}_{1}%
E\cos\phi\right]  \sim\frac{E^{2}}{M}\left(  1-\cos\phi\right)  ,\\
e^{L}\cdot D\cdot k_{1}  &  =\frac{1}{M}\left[  \mathrm{k}_{2}E+\mathrm{k}%
_{1}E\cos\phi\right]  \sim\frac{E^{2}}{M}\left(  1+\cos\phi\right)  ,\\
e^{L}\cdot D\cdot k_{2}  &  =\frac{1}{M}\left[  -\mathrm{k}_{2}E-\mathrm{k}%
_{2}E\right]  \sim-\frac{2E^{2}}{M},
\end{align}
and%
\begin{align}
a_{0}  &  =k_{1}\cdot D\cdot k_{1}=-E^{2}-\mathrm{k}_{1}^{2}\sim-2E^{2},\\
b_{0}  &  =2k_{1}\cdot k_{2}+1=2\left(  E^{2}-\mathrm{k}_{1}\mathrm{k}_{2}%
\cos\phi\right)  +1\sim2E^{2}\left(  1-\cos\phi\right)  ,\\
c_{0}  &  =2k_{1}\cdot D\cdot k_{2}+1=2\left(  E^{2}+\mathrm{k}_{1}%
\mathrm{k}_{2}\cos\phi\right)  +1\sim2E^{2}\left(  1+\cos\phi\right)  .
\end{align}
The scattering amplitude is then calculated to be%

\begin{align}
A_{tensor}  &  =\varepsilon_{T^{n-2q}L^{q},T^{n-2q^{\prime}}L^{q^{\prime}}%
}\int d^{2}z_{1}d^{2}z_{2}\left\langle V_{1}\left(  z_{1},\bar{z}_{1}\right)
V_{2}^{T^{n-2q}L^{q},T^{n-2q^{\prime}}L^{q^{\prime}}}\left(  z_{2},\bar{z}%
_{2}\right)  \right\rangle \nonumber\\
&  =\varepsilon_{T^{n-2q}L^{q},T^{n-2q^{\prime}}L^{q^{\prime}}}\int d^{2}%
z_{1}d^{2}z_{2}\cdot\left\langle e^{ik_{1}X}\left(  z_{1}\right)
e^{ik_{1}\tilde{X}}\left(  \bar{z}_{1}\right)  \right. \nonumber\\
&  \left.  \left(  \partial X^{T}\right)  ^{n-2q}\left(  i\partial^{2}%
X^{L}\right)  ^{q}e^{ik_{2}X}\left(  z_{2}\right)  \left(  \bar{\partial
}\tilde{X}^{T}\right)  ^{n-2q^{\prime}}\left(  i\bar{\partial}^{2}\tilde
{X}^{L}\right)  ^{q^{\prime}}e^{ik_{2}\tilde{X}}\left(  \bar{z}_{2}\right)
\right\rangle \nonumber\\
&  =\left(  -1\right)  ^{q+q^{\prime}}\int d^{2}z_{1}d^{2}z_{2}\left(
z_{1}-\bar{z}_{1}\right)  ^{k_{1}\cdot D\cdot k_{1}}\left(  z_{2}-\bar{z}%
_{2}\right)  ^{k_{2}\cdot D\cdot k_{2}}\left\vert z_{1}-z_{2}\right\vert
^{2k_{1}\cdot k_{2}}\left\vert z_{1}-\bar{z}_{2}\right\vert ^{2k_{1}\cdot
D\cdot k_{2}}\nonumber\\
&  \cdot\left[  \frac{ie^{T}\cdot k_{1}}{z_{1}-z_{2}}+\frac{ie^{T}\cdot D\cdot
k_{1}}{\bar{z}_{1}-z_{2}}+\frac{ie^{T}\cdot D\cdot k_{2}}{\bar{z}_{2}-z_{2}%
}\right]  ^{n-2q}\nonumber\\
&  \cdot\left[  \frac{ie^{T}\cdot D\cdot k_{1}}{z_{1}-\bar{z}_{2}}%
+\frac{ie^{T}\cdot k_{1}}{\bar{z}_{1}-\bar{z}_{2}}+\frac{ie^{T}\cdot D\cdot
k_{2}}{z_{2}-\bar{z}_{2}}\right]  ^{n-2q^{\prime}}\nonumber\\
&  \cdot\left[  \frac{e^{L}\cdot k_{1}}{\left(  z_{1}-z_{2}\right)  ^{2}%
}+\frac{e^{L}\cdot D\cdot k_{1}}{\left(  \bar{z}_{1}-z_{2}\right)  ^{2}}%
+\frac{e^{L}\cdot D\cdot k_{2}}{\left(  \bar{z}_{2}-z_{2}\right)  ^{2}%
}\right]  ^{q}\nonumber\\
&  \cdot\left[  \frac{e^{L}\cdot D\cdot k_{1}}{\left(  z_{1}-\bar{z}%
_{2}\right)  ^{2}}+\frac{e^{L}\cdot k_{1}}{\left(  \bar{z}_{1}-\bar{z}%
_{2}\right)  ^{2}}+\frac{e^{L}\cdot D\cdot k_{2}}{\left(  z_{2}-\bar{z}%
_{2}\right)  ^{2}}\right]  ^{q^{\prime}}.
\end{align}
Set $z_{1}=iy$ and $z_{2}=i$ to fix the $SL(2,R)$ invariance, we have%
\begin{align}
A_{tensor}^{\left(  0\rightarrow1\right)  }  &  =4\left(  2i\right)
^{k_{1}\cdot D\cdot k_{1}+k_{2}\cdot D\cdot k_{2}}\int_{0}^{1}dy\text{
}y^{k_{1}\cdot D\cdot k_{1}}\left(  1-y\right)  ^{2k_{1}\cdot k_{2}+1}\left(
1+y\right)  ^{2k_{1}\cdot D\cdot k_{2}+1}\nonumber\\
&  \cdot\left[  -\frac{e^{T}\cdot k_{1}}{1-y}-\frac{e^{T}\cdot D\cdot k_{1}%
}{1+y}-\frac{e^{T}\cdot D\cdot k_{2}}{2}\right]  ^{n-2q}\nonumber\\
&  \cdot\left[  \frac{e^{T}\cdot D\cdot k_{1}}{1+y}+\frac{e^{T}\cdot k_{1}%
}{1-y}+\frac{e^{T}\cdot D\cdot k_{2}}{2}\right]  ^{n-2q^{\prime}}\nonumber\\
&  \cdot\left[  \frac{e^{L}\cdot k_{1}}{\left(  1-y\right)  ^{2}}+\frac
{e^{L}\cdot D\cdot k_{1}}{\left(  1+y\right)  ^{2}}+\frac{e^{L}\cdot D\cdot
k_{2}}{4}\right]  ^{q}\nonumber\\
&  \cdot\left[  \frac{e^{L}\cdot D\cdot k_{1}}{\left(  1+y\right)  ^{2}}%
+\frac{e^{L}\cdot k_{1}}{\left(  1-y\right)  ^{2}}+\frac{e^{L}\cdot D\cdot
k_{2}}{4}\right]  ^{q^{\prime}}\nonumber\\
&  =\left(  -1\right)  ^{n}4\left(  2i\right)  ^{k_{1}\cdot D\cdot k_{1}%
+k_{2}\cdot D\cdot k_{2}}\left(  E\sin\phi\right)  ^{2n-2\left(  q+q^{\prime
}\right)  }\left(  \frac{E^{2}}{M}\right)  ^{q+q^{\prime}}\nonumber\\
&  \cdot\int_{0}^{1}dy\text{ }y^{k_{1}\cdot D\cdot k_{1}}\left(  1-y\right)
^{2k_{1}\cdot k_{2}+1}\left(  1+y\right)  ^{2k_{1}\cdot D\cdot k_{2}%
+1}\nonumber\\
&  \cdot\left[  \frac{1}{1-y}-\frac{1}{1+y}\right]  ^{2n-2\left(  q+q^{\prime
}\right)  }\cdot\left[  \frac{1-\cos\phi}{\left(  1-y\right)  ^{2}}%
+\frac{1+\cos\phi}{\left(  1+y\right)  ^{2}}-\dfrac{1}{2}\right]
^{q+q^{\prime}}\nonumber\\
&  =\left(  -1\right)  ^{n}4\left(  2i\right)  ^{k_{1}\cdot D\cdot k_{1}%
+k_{2}\cdot D\cdot k_{2}}\left(  2E\sin\phi\right)  ^{2n}\left(  -\dfrac
{1}{8M\sin^{2}\phi}\right)  ^{q+q^{\prime}}\nonumber\\
&  \cdot\sum_{i=0}^{q+q^{\prime}}\sum_{j=0}^{i}\binom{q+q^{\prime}}{i}%
\binom{i}{j}\left(  -2\right)  ^{i}\left(  1-\cos\phi\right)  ^{j}\left(
1+\cos\phi\right)  ^{i-j}\nonumber\\
&  \int_{0}^{1}dy\text{ }y^{k_{1}\cdot D\cdot k_{1}}\left(  1-y\right)
^{2k_{1}\cdot k_{2}+1}\left(  1+y\right)  ^{2k_{1}\cdot D\cdot k_{2}%
+1}\nonumber\\
&  \cdot\left[  \frac{y}{\left(  1-y\right)  \left(  1+y\right)  }\right]
^{2n-2\left(  q+q^{\prime}\right)  }\left[  \frac{1}{1-y}\right]  ^{2j}\left[
\frac{1}{1+y}\right]  ^{2\left(  i-j\right)  }.
\end{align}
Now in the high energy limit, the master formula Eq.(\ref{master}) reduces to%
\begin{align}
I  &  =I_{\left(  0\rightarrow1\right)  }+I_{\left(  1\rightarrow
\infty\right)  }\nonumber\\
&  \simeq2^{-2a-2-N}B\left(  a+1,\frac{b+1}{2}\right)  \left[  \left(
1+\sqrt{\left\vert \dfrac{b}{2a+b}\right\vert }\right)  ^{N}+\left(
1-\sqrt{\left\vert \dfrac{b}{2a+b}\right\vert }\right)  ^{N}\right]
\nonumber\\
&  \equiv2^{-2a-2-N}B\left(  a+1,\frac{b+1}{2}\right)  F_{N},
\end{align}
where%
\begin{align}
n_{a}  &  =2n-2\left(  q+q^{\prime}\right)  ,\\
n_{b}  &  =-2n+2\left(  q+q^{\prime}\right)  -2j,\\
n_{c}  &  =-2n+2\left(  q+q^{\prime}\right)  -2\left(  i-j\right)  ,\\
N  &  =-\left(  2n_{a}+n_{b}+n_{c}\right)  =2i,\\
2a_{0}  &  =k_{1}\cdot D\cdot k_{1}+k_{2}\cdot D\cdot k_{2},\\
F_{N}  &  =\left(  1+\sqrt{\dfrac{1-\cos\phi}{1+\cos\phi}}\right)
^{N}+\left(  1-\sqrt{\dfrac{1-\cos\phi}{1+\cos\phi}}\right)  ^{N}.
\end{align}
The total high energy scattering amplitude can then be calculated to be%
\begin{align}
&  A_{tensor}=A_{tensor}^{\left(  0\rightarrow1\right)  }+A_{tensor}^{\left(
1\rightarrow\infty\right)  }\nonumber\\
&  \simeq\left(  -1\right)  ^{n}4\left(  2i\right)  ^{2a_{0}}\left(
2E\sin\phi\right)  ^{2n}\left(  -\dfrac{1}{8M\sin^{2}\phi}\right)
^{q+q^{\prime}}\nonumber\\
&  \cdot\sum_{i=0}^{q+q^{\prime}}\sum_{j=0}^{i}\binom{q+q^{\prime}}{i}%
\binom{i}{j}\left(  -2\right)  ^{i}\left(  1-\cos\phi\right)  ^{j}\left(
1+\cos\phi\right)  ^{i-j}\cdot2^{-2a-2-N}B\left(  a+1,\frac{b+1}{2}\right)
F_{N}.
\end{align}
The high energy limit of the beta function is%
\begin{equation}
B\left(  a+1,\frac{b+1}{2}\right)  \simeq B\left(  a_{0}+1,\frac{b_{0}+1}%
{2}\right)  \dfrac{a_{0}^{n_{a}}\left(  \dfrac{b_{0}}{2}\right)  ^{n_{b}/2}%
}{\left(  a_{0}+\dfrac{b_{0}}{2}\right)  ^{n_{a}+n_{b}/2}}.
\end{equation}
Finally we get the scattering amplitudes at mass level $M^{2}=2(n-1)$%
\begin{align}
A_{tensor}  &  =A_{tensor}^{\left(  0\rightarrow1\right)  }+A_{tensor}%
^{\left(  1\rightarrow\infty\right)  }\nonumber\\
&  =\left(  -1\right)  ^{a_{0}}E^{2n}\left(  -\dfrac{1}{2M}\right)
^{q+q^{\prime}}B\left(  a_{0}+1,\frac{b_{0}+1}{2}\right) \nonumber\\
&  \cdot\sum_{i=0}^{q+q^{\prime}}\sum_{j=0}^{i}\binom{q+q^{\prime}}{i}%
\binom{i}{j}\left(  -2\right)  ^{-i}\left(  1+\cos\phi\right)  ^{i}\left(
-1\right)  ^{j}F_{N}\nonumber\\
&  =2\left(  -1\right)  ^{a_{0}}E^{2n}\left(  -\dfrac{1}{2M}\right)
^{q+q^{\prime}}B\left(  a_{0}+1,\frac{b_{0}+1}{2}\right)  ,
\end{align}
where the high energy limit of $B(a_{0}+1,\frac{b_{0}+1}{2})$ is independent
of $q+q^{\prime}$. We thus have explicitly shown that there is only one
independent high energy scattering amplitude at each fixed mass level. It is a
remarkable result that the ratios $\left(  -\dfrac{1}{2M}\right)
^{q+q^{\prime}}$for different high energy scattering amplitudes at each fixed
mass level is consistent with Eq.(\ref{11}) for the scattering without D-brane
as expected. For the two tensor scatterings, the ratios are $\Sigma_{i=1}%
^{2}\left(  -\dfrac{1}{2M_{i}}\right)  ^{(q_{i}+q_{i}^{\prime})}.$

\section{Conclusion}

In this paper, we have shown that the $(0\rightarrow1)$ and $(1\rightarrow
\infty)$ channels scattering amplitudes of an aribitrary closed string state
scattered from D-brane to another arbitrary closed string state can be
expressed in terms of the hypergeometric function $_{2}F_{1}$, which in turn
can be reduced to the usual beta function. We have noted that, mathematically,
the reduction of hypergeometric function $_{2}F_{1}$ to the beta function in
the string scattering amplitudes is mainly due to the Kummer condition
Eq.(2.22). Physically, this condition is realized in the equation of momentum
conservation on the D-brane Eq.(2.17). After summing up the $(0\rightarrow1)$
and $(1\rightarrow\infty)$ channels, we discover that \textit{all} the
scattering amplitudes can be expressed in terms of the generalized
hypergeometric function $_{3}F_{2}$ with special arguments, which terminates
to a finite sum and, as a result, the whole scattering amplitudes consistently
reduce to the usual beta function. Our results suggest the interesting
relation between string scattering amplitudes and the (generalized)
hypergeometric functions both physically and mathematically.

Finally, we explicitly calculate high-energy limit of a series of the above
scattering amplitudes for arbitrary mass levels, and derive the linear
relations among them for the case of D-particle. Since the calculation of
decoupling of high energy ZNS in D-brane scatterings remains the same as the
case of scatterings without D-brane, the ratios of these high-energy
scattering amplitudes at each fixed mass level are found to be consistent with
the calculations of our previous works without D-brane. Presumably, this
result can be generalized to the case of general D$p$-brane except D-instanton
and Domain-wall \cite{Wall}.

\section{Acknowledgments}

This work is supported in part by the National Science Council and National
Center of Theoretical Science, Taiwan, R.O.C.

\appendix%

\setcounter{equation}{0}
\renewcommand{\theequation}{\thesection.\arabic{equation}}%

\section{A Brief Review of $_{2}F_{1}$ and $_{3}F_{2}$}

In this appendix, we review the definitions and some formulas of
hypergeometric function $_{2}F_{1}$ and generalized hypergeometric function
$_{3}F_{2}$ which we used in the text. hypergeometric functions form an
important class of special functions. Many elementary special functions are
special cases of $_{2}F_{1}$. The hypergeometric function $_{2}F_{1}$ is
defined to be ($\alpha,\beta,\gamma$ constant)%

\begin{align}
_{2}F_{1}(\alpha,\beta,\gamma;x)  &  =1+\frac{\alpha\beta}{\gamma}\frac{x}%
{1!}+\frac{\alpha(\alpha+1)\beta(\beta+1)}{\gamma(\gamma+1)}\frac{x^{2}}%
{2!}+\cdot\cdot\cdot=\sum_{n=0}^{\infty}\frac{(\alpha)_{n}(\beta)_{n}}%
{(\gamma)_{n}}\frac{x^{n}}{n!}\nonumber\\
&  =\frac{\Gamma(\gamma)}{\Gamma(\alpha)\Gamma(\beta)}\sum_{n=0}^{\infty}%
\frac{\Gamma(\alpha+n)\Gamma(\beta+n)}{\Gamma(\gamma+n)}\frac{x^{n}}{n!}%
\end{align}
where%
\begin{equation}
(\alpha)_{0}=1,(\alpha)_{n}=\alpha(\alpha+1)(\alpha+2)\cdot\cdot\cdot
(\alpha+n-1)=\frac{(\alpha+n-1)!}{(\alpha-1)!}.
\end{equation}
The hypergeometric function $_{2}F_{1}$ is a solution, at the singular point
$x=0$ with indicial root $r=0$, of the Gauss's hypergeometric differential
equation%
\begin{equation}
x(1-x)u^{\prime\prime}+\left[  \gamma-(\alpha+\beta+1)\right]  u^{\prime
}-\alpha\beta u=0, \label{DE}%
\end{equation}
which contains three regular singularities $x=0,1,\infty$. The second solution
of Eq. (\ref{DE}) with indicial root $r=1-\gamma$ can be expressed in terms of
$_{2}F_{1}$ as following ($\gamma\neq$ integer)%
\begin{equation}
u_{2}(x)=x^{1-\gamma}2F_{1}(\alpha-\gamma+1,\beta-\gamma+1,2-\gamma,x).
\end{equation}
Other solutions of Eq. (\ref{DE}), which corresponds to singularities
$x=1,\infty$, can also be expressed in terms of the hypergeometric function
$_{2}F_{1}$. The following identity
\begin{equation}
_{2}F_{1}(\alpha,\beta,\gamma;x)=(1-x)^{\gamma-\alpha-\beta}\text{ }_{2}%
F_{1}(\gamma-\alpha,\gamma-\beta,\gamma;x),
\end{equation}
which we used in the text can then be derived.

$_{2}F_{1\text{ }}$has an intergal representation%
\begin{equation}
_{2}F_{1}(\alpha,\beta,\gamma;x)=\frac{\Gamma(\gamma)}{\Gamma(\beta
)\Gamma(\gamma-\beta)}\int_{0}^{1}dy\text{ }y^{\beta-1}\left(  1-y\right)
^{\gamma-\beta-1}\left(  1-yx\right)  ^{-\alpha}, \label{Key}%
\end{equation}
which can be used to do analytic continuation. Equation (\ref{Key}) with
$x=-1$ was repeatedly used in the text in our calculations of string
scattering amplitudes with D-brane.

There exists interesting relations among hypergeometric function $_{2}F_{1}$
with different arguments%
\begin{equation}
x^{-p}(1-x)^{-q}2F_{1}(\alpha,\beta,\gamma;x)=t^{-p^{\prime}}(1-t)^{-q^{\prime
}}\text{ }_{2}F_{1}(\alpha^{\prime},\beta^{\prime},\gamma^{\prime};t),
\end{equation}
where $x=\varphi(t)$ is an algebraic function with degree up to six. As an
example, the quadratic transformation formula%
\begin{equation}
_{2}F_{1}(\alpha,\beta,1+\alpha-\beta;x)=(1-x)^{-\alpha}\text{ }_{2}%
F_{1}\left(  \frac{\alpha}{2},\frac{1+\alpha-2\beta}{2},1+\alpha-\beta
;\frac{-4x}{(1-x)^{2}}\right)  ,
\end{equation}
can be used to derive the Kummer's relation%
\begin{equation}
_{2}F_{1}(\alpha,\beta,1+\alpha-\beta,-1)=\frac{\Gamma(1+\alpha-\beta
)\Gamma(1+\frac{\alpha}{2})}{\Gamma(1+\alpha)\Gamma(1+\frac{\alpha}{2}-\beta
)},
\end{equation}
which is crucial to reduce the scattering amplitudes of string from D-brane to
the usual beta function.

In summing up the $(0\rightarrow1)$ and $(1\rightarrow\infty)$ channel
scattering amplitudes, we have used the master formula%

\begin{align}
I  &  =I_{\left(  0\rightarrow1\right)  }+I_{\left(  1\rightarrow
\infty\right)  }\nonumber\\
&  =2^{-2a-1-N}\int_{0}^{1}dt\text{ }t^{b}\left(  1-t\right)  ^{a}\left(
1+t\right)  ^{a}\left[  \left(  1+t\right)  ^{N}+\left(  1-t\right)
^{N}\right] \nonumber\\
&  =2^{-2a-1-N}\sum_{n=0}^{\left[  \frac{N}{2}\right]  }\binom{N}{2n}\cdot
B\left(  a+1,\frac{b+1}{2}+n\right) \nonumber\\
&  =2^{-2a-1-N}\cdot B\left(  a+1,\frac{b+1}{2}\right)  \cdot\text{ }_{3}%
F_{2}\left(  \frac{b+1}{2},-\left[  \frac{N}{2}\right]  ,\dfrac{1}{2}-\left[
\frac{N}{2}\right]  ;a+\frac{b+3}{2},\dfrac{1}{2};1\right)  . \label{3F2}%
\end{align}
In Equation (\ref{3F2}), $B$ is the beta function and $_{3}F_{2}$ is the
generalized hypergeometric function, which is defined to be%
\begin{equation}
_{3}F_{2}(\alpha_{1},\alpha_{2},\alpha_{3};\gamma_{1},\gamma_{2};x)=\sum
_{n=0}^{\infty}\frac{(\alpha_{1})_{n}(\alpha_{2})_{n}(\alpha_{3})_{n}}%
{(\gamma_{1})_{n}(\gamma_{2})_{n}}\frac{x^{n}}{n!}.
\end{equation}
For those arguments of $_{3}F_{2}$ in Eq. (\ref{3F2}), the series of the
generalized hypergeometric function $_{3}F_{2}$ terminates to a finite sum.
For example,%
\begin{align}
N  &  =0:\text{ }_{3}F_{2}=1,\nonumber\\
N  &  =1:\text{ }_{3}F_{2}=1,\nonumber\\
N  &  =2:\text{ }_{3}F_{2}=\dfrac{a+b+2}{a+\frac{b+3}{2}},\nonumber\\
N  &  =3:\text{ }_{3}F_{2}=\dfrac{a+2b+3}{a+\frac{b+3}{2}},\nonumber\\
N  &  =4:\text{ }_{3}F_{2}=\dfrac{a^{2}+4ab+2b^{2}+7a+12b+12}{\left(
a+\frac{b+3}{2}\right)  \left(  a+\frac{b+5}{2}\right)  },\nonumber\\
N  &  =5:\text{ }_{3}F_{2}=\dfrac{a^{2}+6ab+9b^{2}+4a+22b+20}{\left(
a+\frac{b+3}{2}\right)  \left(  a+\frac{b+5}{2}\right)  }.
\end{align}


\begin{thebibliography}{99}                                                                                               %


\bibitem {Massive}F.Liu, Phys. Rev. D38,1334 (1988). J.~C.~Lee,
Prog.\ Theor.\ Phys.\ \textbf{91}, 353 (1994); Phys.\ Lett.\ B \textbf{337},
69 (1994);

\bibitem {GM}D.~J.~Gross and P.~F.~Mende,
Phys.\ Lett.\ B \textbf{197}, 129 (1987);
Nucl.\ Phys.\ B \textbf{303}, 407 (1988).

\bibitem {Gross}D.~J.~Gross,
Phys.\ Rev.\ Lett.\ \textbf{60}, 1229 (1988); Phil.\ Trans.\ R. Soc. Lond.
A329, 401 (1989).

\bibitem {GrossManes}D.~J.~Gross and J.~L.~Manes,
Nucl.\ Phys.\ B \textbf{326}, 73 (1989). See section 6 for details.

\bibitem {ChanLee1}C.~T.~Chan and J.~C.~Lee,
Phys.\ Lett.\ B \textbf{611}, 193 (2005).
J.~C.~Lee,
[arXiv:hep-th/0303012].

\bibitem {ChanLee2}C.~T.~Chan and J.~C.~Lee,
Nucl.\ Phys.\ B \textbf{690}, 3 (2004).


\bibitem {CHL}C.~T.~Chan, P.~M.~Ho and J.~C.~Lee,
Nucl.\ Phys.\ B \textbf{708}, 99 (2005).


\bibitem {CHLTY}C.~T.~Chan, P.~M.~Ho, J.~C.~Lee, S.~Teraguchi and Y.~Yang,
Nucl.\ Phys.\ B \textbf{725}, 352 (2005).


\bibitem {PRL}C.~T.~Chan, P.~M.~Ho, J.~C.~Lee, S.~Teraguchi and Y.~Yang,
Phys. Rev. Lett. 96 (2006) 171601.

\bibitem {paperB}C.~T.~Chan, P.~M.~Ho, J.~C.~Lee, S.~Teraguchi and Y.~Yang,
Nucl.\ Phys.\ B \textbf{749}, 266 (2006).
\textquotedblleft Comments on the high energy limit of bosonic open string
theory,\textquotedblright\ [arXiv:hep-th/0509009].

\bibitem {susy}C.~T.~Chan, J.~C.~Lee and Y.~Yang,
Nucl.\ Phys.\ B \textbf{738}, 93 (2006).


\bibitem {Closed}C.~T.~Chan, J.~C.~Lee and Y.~Yang,
Nucl.\ Phys.\ B \textbf{749}, 280 (2006).


\bibitem {HL}Pei-Ming Ho, Xue-Yan Lin, Phys.Rev. D73 (2006) 126007.

\bibitem {ZNS1}J.~C.~Lee,
Phys.\ Lett.\ B \textbf{241}, 336 (1990); Phys.\ Rev.\ Lett.\ \textbf{64},
1636 (1990). J.~C.~Lee and B.~Ovrut,
Nucl.\ Phys.\ B \textbf{336}, 222 (1990), Phys.\ Lett.\ B \textbf{326}, 79
(1994).


\bibitem {ZNS3}T.~D.~Chung and J.~C.~Lee,
Phys.\ Lett.\ B \textbf{350}, 22 (1995).
Z.\ Phys.\ C \textbf{75}, 555 (1997).
J.~C.~Lee,
Eur.\ Phys.\ J.\ C \textbf{1}, 739 (1998).


\bibitem {ZNS2}H.~C.~Kao and J.~C.~Lee,
Phys.\ Rev.\ D \textbf{67}, 086003 (2003).
. C.~T.~Chan, J.~C.~Lee and Y.~Yang,
Phys.\ Rev.\ D \textbf{71}, 086005 (2005)


\bibitem {Klebanov}For a review, see A. Hashimoto and I.R. Klebanov,
"Scattering of strings from D-branes" hep-th/9611214 and references therein.
M.R. Garousi and R.C. Myers, "Superstring scattering from D-Branes" Nucl.
Phys. B475 (1996) 193, hep-th/9603194. I.R. Klebanov and L. Thorlacius, Phys.
Lett. B371,51 (1996). J.L.F. Barbon, Phys. Lett. B382, 60 (1996). C. Bachas
and B. Pioline, JHEP9912, 004 (1999). S. Hirano and Y. Kazama, Nucl. Phys.
B499, 495 (1997).

\bibitem {GSW}M.~B.~Green, J.~H.~Schwarz and E.~Witten, \textquotedblleft
Superstring Theory. Vol. 1,\textquotedblright\ Cambridge University Press 1987.

\bibitem {Wall}C.~T.~Chan, J.~C.~Lee and Y.~Yang, "Power-law Behavior of
Strings Scattered from Domain-wall and Breakdown of Their High Energy Linear
Relations", hep-th/0610219.
\end{thebibliography}
\end{document}